\documentclass[preprint, 12pt]{elsarticle}

\usepackage{lineno,hyperref}
\usepackage{bmpsize}
\usepackage{graphicx}
\usepackage{multirow} 
\usepackage{tabularx} 
\usepackage{dcolumn}
\usepackage{bm}
\usepackage{color}
\usepackage[T1]{fontenc}
\modulolinenumbers[5]
\usepackage{picins}

\usepackage{color}

\newcommand{\jj}{\textcolor{black}}
\newcommand{\jjNew}{\textcolor{black}}
\newcommand{\vp}{\textcolor{black}}
\newcommand{\pa}{\textcolor{black}}
\newcommand{\paNew}{\textcolor{black}}

\usepackage{numcompress}

\begin{document}

\begin{frontmatter}

\title{Minimal Perceptrons for Memorizing Complex Patterns}

\author[apctp,usc]{Marissa Pastor\fnref{fn1}}
\fntext[fn1]{These authors contributed equally to this work}
\author[apctp,postech]{Juyong Song\fnref{fn1}}
\author[apctp,qbu]{Danh-Tai Hoang}
\author[apctp,postech]{Junghyo Jo\corref{mycorrespondingauthor}}
\cortext[mycorrespondingauthor]{Corresponding author}
\ead{jojunghyo@apctp.org}

\address[apctp]{Asia Pacific Center for Theoretical Physics, Pohang, Korea}
\address[usc]{Department of Physics, University of San Carlos Talamban, Cebu, Philippines}
\address[postech]{Department of Physics, Pohang University of Science and Technology, Pohang, Korea}
\address[qbu]{Department of Science, Quang Binh University, Dong Hoi, Vietnam}

\begin{abstract}
Feedforward neural networks have been investigated to understand learning and memory, as well as applied to numerous practical problems in pattern classification. \jjNew{It is a rule of thumb that more complex tasks require larger networks. However, the design of optimal network architectures for specific tasks is still an unsolved fundamental problem. In this study, we consider three-layered neural networks for memorizing binary patterns. We developed a new complexity measure of binary patterns, and estimated the minimal network size for memorizing them as a function of their complexity. We formulated the minimal network size for regular, random, and complex patterns. In particular, the minimal size for complex patterns, which are neither ordered nor disordered, was predicted by measuring their Hamming distances from known ordered patterns.} Our predictions agreed with simulations based on the back-propagation algorithm.
\end{abstract}


\end{frontmatter}


\section{Introduction}
\label{sec:Intro}
 Neural \vp{signaling in} synaptic networks motivated \vp{the early study of} artificial neural networks, \vp{to recapitulate} the learning capability of the brain.
 Their utility has expanded from \vp{that} inception to \vp{serving as} alternatives to conventional computers 
 for  input/output processing \vp{or as exemplars of} parallel distributed processing, 
and \vp{they have} been successfully applied \vp{to} pattern classification~\citep{GOu2007, Monterola2005} and memory storage~\citep{Hopfield1982, Abbott1987, Shiino1993, Watkin1993}. 
Standard implementations of neural networks 
map inputs $\bm{x}^\mu$ \vp{to}  outputs ${z}^\mu$ \vp{through intermediate processing layers}.
The $M$ input/output pairs, $\xi^\mu=(\bm{x}^\mu, {z}^\mu)$, form a {\it pattern}, $\bm{\xi}=\{ \xi^1, \xi^2, ..., \xi^M \}$.
This input/output mapping can be achieved in two different ways:
The neural network can either (i) learn the underlying rule for the mapping from some training pairs,
or (ii) memorize the whole pattern of input/output pairs and retrieve the stored outputs in response to given inputs.
In either \vp{way}, it is a major impediment that the required complexity
of network architectures for learning or memorizing certain patterns is, in general, \vp{unknown}. 
In this paper, we focus on memorizing patterns.

Designing the optimal network architecture has been known as an NP (Non-deterministic Polynomial-time) problem
that requires computationally expensive search techniques and optimization~\citep{Blum1992, Murata1994, Hirose1991, Hunter2012}.
Indeed, most \vp{attempts} use empirical approaches and proceed by scanning over different network configurations 
while utilizing  incremental~\citep{Kwok1997} and/or pruning algorithms~\citep{Reed1993, Castellano1997}.
Simple networks {may lead to insufficient memory and poor generalization},
while complex networks lead to \vp{poor predictive performance} by overestimating each element in patterns~\citep{Watkin1993}.
The required complexity of networks generally depends on the complexity of patterns for memorizing.
Therefore, if the complexity of patterns and networks could be quantified, 
the optimal network architecture could be systematically designed.

Two popular complexity measures for patterns are 
Shannon entropy, the degree of uncertainty for describing a pattern~\citep{Shannon1948, Silva2015, Larrondo2005},
and Kolmogorov complexity, the length of the shortest computer program for generating a pattern~\citep{Cover2006}.
However, the uncertainty or probability of each element in a pattern is \vp{unknown} and the 
algorithmic complexity is \vp{itself difficult to compute}.
These difficulties \vp{suggest the need for} a metric to quantify the \vp{practical} complexity of patterns \vp{relevant for perceptrons}.
Here we propose a simple complexity index and relate it to the minimal network size for memorizing patterns.

This paper is organized as follows: 
We introduce the mathematical description of \pa{our} feedforward neural network in Sec.~\ref{sec:NN},
and a storage problem of binary patterns of different complexities in Sec.~\ref{sec:complexity}.
Next, we estimate minimal network sizes for storing regular binary patterns in Sec.~\ref{sec:parity}
and random binary patterns in Sec.~\ref{sec:random}, and compare them to simulation results.
Then, we generalize the complexity formulation to estimate minimal network size for storing complex binary patterns in Sec.~\ref{sec:complex}.
Finally, we summarize the paper in Sec.~\ref{sec:conclusion}.

\section{Neural network}
\label{sec:NN}
\begin{figure} 
\begin{center}
\includegraphics[width=100mm]{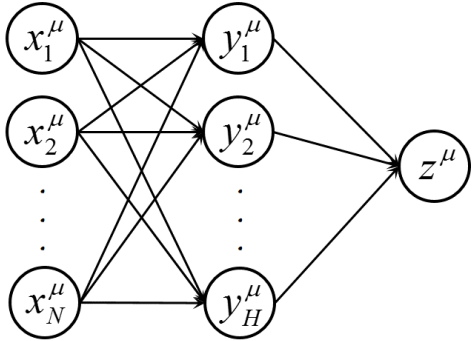}
\caption{ 
Three-layer feedforward neural network.
\label{fig1}
}
\end{center}
\end{figure}

We study a three-layer feedforward neural network as shown in Fig.~\ref{fig1}.
This simple network architecture \vp{is} successful at solving pattern recognition problems~\citep{LeCun1998, Haykin1999}.
In addition, the universal approximation theorem proves that the three-layer network \vp{suffices} to approximate any continuous function, 
${z}^\mu = {f}(\bm{x}^\mu)$~\citep{Cybenko1989}. 
For simplicity, we consider $N$-dimensional vectors of binary inputs $\bm{x}^\mu = (x^{\mu}_1, x^{\mu}_2, ..., x^{\mu}_N)$ and
scalar binary outputs $z^\mu$. 
One pattern $\bm \xi$ represents $2^N$ pairs of $(\bm x^\mu, z^\mu)$, 
because each component in the $N$-dimensional input vector takes \vp{values}  $x^\mu_i =0$ or $1$. 
The input/output mapping \vp{requires} $N$ input nodes and a single output node. 
In the feedforward three-layer network, an input $\bm{x}^\mu$ \vp{is transformed} into 
the activities $\bm{y}^\mu=\{y^{\mu}_1, y^{\mu}_2, ..., y^{\mu}_H\}$ of $H$ hidden nodes:
\begin{equation} \label{eq:y}
y^{\mu}_j= \sigma \Big(\sum_{i=1}^{N} w_{ji} x^{\mu}_i - w_{j0} \Big),
\end{equation}
where $w_{ji}$ is the connection weight from the $i$th input node to the $j$th hidden node,
and $w_{j0}$ is the bias of the $j$th hidden node. 
\vp{With these definitions,} the $j$th hidden node is activated when the integrated input signal
$\sum_i w_{ji} x^{\mu}_i$ exceeds the bias $w_{j0}$ through the sigmoidal activation function, $\sigma(a)=1/(1+e^{-a})$.
The transformation from the hidden layer to \pa{the} output layer follows the same rule:
\begin{equation} \label{eq:z}
z^{\mu}= \sigma \Big(\sum_{j=1}^{H} v_{j} y^{\mu}_j - v_0 \Big),
\end{equation}
where $v_{j}$ is the connection weight from the $j$th hidden node to \pa{the} output node,
and $v_0$ is the bias of \pa{the} output node.
Successful storage of the pattern $\bm{\xi}$ represents
the \vp{correct} input/output transformation through the feedforward network
with \vp{appropriate} parameters ($w_{ji}, v_j$), $i\in \{0, 1, ..., N\}$ and $j \in \{0, 1, ..., H\}$.
The required minimum network size (i.e., number of hidden nodes, $H$) 
for the \vp{successful} storage of a certain pattern is \vp{the} key question \vp{we address}.

\section{Pattern complexity}
\label{sec:complexity}
\begin{figure}
\begin{center}
\includegraphics[width=100mm]{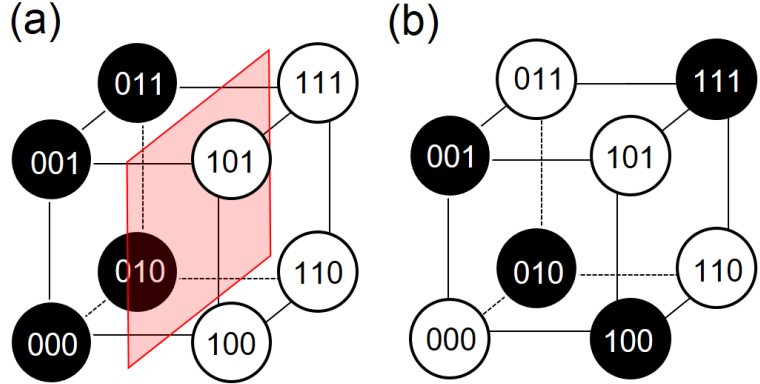}
\caption{(Color online) Binary patterns.
(a) A simple pattern and (b) a complex pattern for an input dimension $N=3$.
A hyperplane (red) separates two groups of black and white elements.
\label{fig2}
}
\end{center}
\end{figure}

\vp{An example of the problem} of  binary pattern storage is a dichotomy problem \vp{on} a binary $N$-cube.
\vp{For a linearly separable problem,} the transformation from inputs to the $j$th hidden node corresponds to
the dichotomy $\{Y^+_j, Y^-_j \}$ of the elements $\bm{x}^{\mu}$
above and below an $(N-1)$-dimensional hyperplane, 
$\sum_{i=1}^N w_{ji} x_i^\mu - w_{j0} = 0$.
\vp{Thus, for} a simple binary pattern, one hyperplane is sufficient to dichotomize
$\xi^\mu$ with respect to $z^\mu=0$ and $z^\mu=1$ elements (Fig.~\ref{fig2}a).
This means that one hidden node is sufficient to store the simple pattern.
For more complex patterns (Fig.~\ref{fig2}b), however, additional hidden nodes (or hyperplanes) 
and  processing from the hidden layer to \pa{the} output layer are necessary.
The complexity of binary patterns is \vp{lower}
as neighboring elements are homogeneous in the binary $N$-cube. 
This observation \vp{suggests} a simple complexity index:
\begin{equation} \label{eq:complexity}
K=\frac{1}{M} \sum_{\mu=1}^M k^\mu,
\end{equation}
where $k^\mu$ is the number of different neighboring elements for an element $\xi^\mu$. 
Considering the binary vectors $\bm{x}^{\mu'}$, one-bit different from the vector $\bm{x}^\mu=(x^\mu_1, x^\mu_2, ..., x^\mu_N)$,
as the neighborhood set $\Lambda_\mu$ of $\bm{x}^\mu$,
the individual complexity index is defined as $k^\mu = \sum_{\mu' \in \Lambda_\mu} (1-\delta_{z^\mu, z^{\mu'}})$,
using the Kronecker delta function $\delta_{z, z'} = 1$ for $z=z'$, and $0$ otherwise.

\section{Minimal network for regular and random patterns}
\subsection{Regular patterns}
\label{sec:parity}
The minimum number of hidden nodes \vp{required} for memorizing some regular patterns is known.
Parity patterns, $z^\mu = \bmod(\sum_{i=1}^N x^\mu_i, 2)$,
require $H=N/2+1$ and $H=(N+1)/2$ hidden nodes
for even and odd $N$ input dimensions, respectively~\citep{Setiono1997, Sontag1992}.
Parity patterns have the complexity index $K=N$,
because every neighbor of an element has different outputs despite one-bit difference in their inputs.
One may generate simpler regular patterns by introducing pseudo bits
that \vp{have no effect on the}  output.
When $n$ pseudo-bits are introduced \vp{amongst the} input bits
\pa{such that the effective inputs become ($N-n$) bits,
this ($N$, $n$) pseudo-parity pattern requires \vp{fewer} hidden nodes for memorizing the pattern.}
The pseudo pattern has a reduced complexity index $K=N-n$.
For \vp{such} regular patterns, the minimum number of hidden nodes is generally formulated as
\begin{equation} \label{eq:H1}
H = \frac{K}{2} + 1
\end{equation}
for even $N$ \pa{and $H=(K+1)/2$ for odd $N$.}

\subsection{Random patterns}
\label{sec:random}
The complexity index $K$ measures local complexities in patterns.
Thus it is \vp{insufficient if we want} to capture global order \vp{underlying} patterns.
Parity patterns look less complex than random patterns,
although they have the highest complexity index. 
\vp{Such patterns} have a strong order or rule that
makes numerous elements, having \pa{identical values for $\sum_{i=1}^N (-1)^i x^\mu_i$,} redundant~\citep{Setiono1997}.
Random patterns, lacking any order, do not have redundant elements for memorizing.
Therefore, it is \vp{to be} expected that random patterns require \vp{the} maximum number of hidden nodes
for memorizing, given $K$.
Although the storage of random patterns itself may be practically useless,
it determines the upper bound of $H$ for the storage of binary patterns.

\begin{figure}
\begin{center}
\includegraphics[width=100mm]{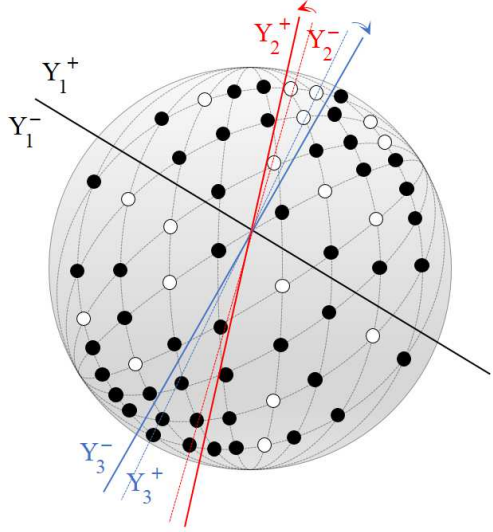}
\caption{(Color online) Schematic diagram for the storage of binary random patterns.
A reference hyperplane $Y_1$ is introduced for the dichotomy of black and white elements
\jj{in $Y_1^+$ and $Y_1^-$, respectively.
Additional hyperplanes $Y_2$ and $Y_3$ isolate impurities of white elements in $Y_1^+$
and black elements in $Y_1^-$.}
\label{fig3}
}
\end{center}
\end{figure}

Here we propose one strategy for memorizing random patterns.
Suppose that a random binary pattern has \vp{a} total \vp{of} $M(=2^N)$ elements 
in which $pM$ elements \vp{are black} ($z^\mu=1$),
and $(1-p)M$ elements \vp{are white} ($z^\mu=0$).
First, we introduce a reference hyperplane $Y_1$ dividing the $N$-cube into two regions,
$Y_1^+$ and $Y_1^-$ above and below the hyperplane,
that contain $pM$ and $(1-p)M$ elements, respectively (Fig.~\ref{fig3}).
Then, \vp{the} $Y_1^+$ region has $(1-p)pM$ {\it impurities} of white elements,
while \vp{the} $Y_1^-$ region has $p(1-p)M$ impurities of black elements.

\jj{Next, by introducing a pair of hyperplanes, $Y_2$ and $Y_3$,
we can isolate clustered impurities above/below the reference hyperplane.
As shown in Table~\ref{tab:table1},
the paired hyperplanes can selectively correct the outputs of the impurities
without perturbing the outputs of the other elements.
In principle, we can define the paired hyperplanes from clustered impurities.
Suppose we choose ($N+1$) impurities $\xi^\mu$ to define a $Y_2$ hyperplane.
These impurities imply  ($N+1$) linear equations, $\sum_{i=1}^N w_{2i} x^\mu_i - w_{20} =0$,
which can also be described as a matrix equation, $\bm{X}_2 \cdot \bm{w}_2 = \bm{0}$.
Similarly we can define a $Y_3$ hyperplane with $\bm{X}_3 \cdot \bm{w}_3 = \bm{0}$ 
by choosing another ($N+1$) impurities which are located sufficiently close to the $Y_2$ hyperplane.
Then we can place those $2(N+1)$ impurities into a small interspace between the $Y_2$ and $Y_3$ hyperplanes
by slightly rotating the two hyperplanes in opposite directions with respect to their intersection (Fig.~\ref{fig3}):
$\bm{X}_2 \cdot \bm{w}_2 = \bm{\epsilon}_2$ and $\bm{X}_3 \cdot \bm{w}_3 = \bm{\epsilon}_3$.
Here we can uniquely determine the paired hyperplanes for isolating $2(N+1)$ impurities, 
if both matrices $\bm{X}_2$ and $\bm{X}_3$ have rank $(N+1)$.
This implies that one additional hyperplane can isolate ($N+1$) impurities.
To isolate a total of $2p(1-p)M$ impurities, we need $2p(1-p)M/(N+1)$ hyperplanes.
Therefore, the minimum number of hidden nodes for memorizing random patterns
corresponds to the total number of the impurity hyperplanes in addition to the single reference hyperplane:
}
\begin{equation} \label{eq:H2a}
H=\frac{2^N K}{N(N+1)} + 1,
\end{equation}
where the complexity index $K=2p(1-p)N$ is derived from Eq.~(\ref{eq:complexity}).
The random pattern has $pM$ black elements with $k^\mu=(1-p)N$,
and $(1-p)M$ white elements with $k^\mu=pN$ \vp{on average}.

\begin{table}
\caption{\label{tab:table1} Dichotomy of random binary patterns.
The weight parameters are chosen as $v_j=2$ 
and the bias as $v_0=3$. 
}
\begin{tabularx}{\textwidth}{XX}
\begin{tabular}{c|ccc|c}
\hline\hline
 &$y_1$ & $y_2$ & $y_3$ & $\sum_j v_j y_j - v_0$ \\
 \hline
$Y_1^+ Y_2^+ Y_3^-$ & 1 & 1 & 0 & $+1$\\
$Y_1^+ Y_2^- Y_3^+$ & 1 & 0 & 1 & $+1$\\
$Y_1^+ Y_2^- Y_3^-$ & 1 & 0 & 0 & $-1$\\
\hline
$Y_1^- Y_2^+ Y_3^-$ & 0 & 1 & 0 & $-1$\\
$Y_1^- Y_2^- Y_3^+$ & 0 & 0 & 1 & $-1$\\
$Y_1^- Y_2^+ Y_3^+$ & 0 & 1 & 1 & $+1$\\
\hline\hline
\end{tabular}
\end{tabularx}
\end{table}

The minimum number $H$ for random patterns can be further reduced if we use 
unexpected \vp{relations} between the conjugate elements $\xi^\mu$ and $\xi^{\mu^*}$
linked by their inputs as $\bm{x}^\mu + \bm{x}^{\mu^*} =  (1,1,..., 1)$.
Suppose that we choose $N$ impurity elements \vp{for} which $N$ conjugate elements
have opposite colors ($z^\mu \neq z^{\mu^*}$).
If the conjugate elements are also impurities located in the opposite side of the reference hyperplane $Y_1$, 
one hyperplane is sufficient to
define $2N$ impurity elements. 
These conjugate impurity elements have $2N$ equations: $\sum_{i=1}^N w_{ji} x^\mu_i - w_{j0} = 0$
and $\sum_{i=1}^N w_{ji} x^{\mu^*}_i - w_{j0} = 0$.
However, the conjugate symmetry reduces $2N$ equations \vp{to}
$(N+1)$ equations: $\sum_{i=1}^N w_{ji} x^\mu_i - w_{j0} = 0$
and $\sum_{i=1}^N w_{ji}  - 2w_{j0} = 0$.
\vp{Thus}, the $(N+1)$ variables  $w_{ji}$ can be uniquely determined. 

This is one scenario out of four possibilities depending on \vp{the} color and location of the conjugate elements:
(i) different color and opposite side to $Y_1$;
(ii) same color and same side;
(iii) different color and same side; and (iv) same color and different side.
As explained for \pa{Case (i), Case (ii)} can also define $2N$ impurities \pa{using} one hyperplane.
However, \pa{Cases} (iii) and (iv) cannot use the conjugate symmetry,
because the conjugate elements are not impurities in $Y_1$.
\pa{For these cases, we} can define only ($N+1$) impurities with one hyperplane.
Among $m (\equiv 2p(1-p)M)$ impurities, 
$2p(1-p) m$ impurities correspond to \pa{Cases} (i) and (ii), 
while $[p^2+(1-p)^2] m$ impurities correspond to \pa{Cases} (iii) and (iv).
Thus the minimum number of hyperplanes is
$H=2p(1-p) m/2N + [p^2+(1-p)^2] m/(N+1) + 1$.
This gives \vp{a} second-order correction \vp{to} Eq.~(\ref{eq:H2a}):
\begin{equation} \label{eq:H2b}
H=\frac{2^N K}{N(N+1)} \Big[1 - \frac{(N-1)K}{2N^2} \Big] + 1.
\end{equation}

The estimated $H$ from Eqs.~(\ref{eq:H1}) and (\ref{eq:H2b}) for regular and random patterns, respectively, are tested  by simulating supervised three-layer feedforward neural networks. Once a network with a certain number of hidden nodes is given, inputs $\bm{x}^\mu$ propagate to an output ${\tilde z}^\mu$ through the hidden layer, \pa{as described in Eqs.~(\ref{eq:y}) and (\ref{eq:z}).} 
Then we measure the mismatches between the computed outputs ${\tilde z}^\mu$ and the true outputs ${z}^\mu$ as an error, $E = 1/2 \sum_{\mu=1}^M (z^\mu - \tilde{z}^\mu)^2$.
Here we optimize the weight and bias parameters with the back-propagation \paNew{(BP)} algorithm~\citep{Rumelhart1986}, a \pa{gradient-descent} method that updates parameters \vp{according to the change needed} for decreasing the error $E$. For example, the update of weights $w_{ji}$ follows $\Delta w_{ji} = - \alpha \partial E / \partial w_{ji}$, in which the learning rate is optimized as $\alpha=0.35$ for \pa{fine, but not slow, searches} for the error-landscape minima. \pa{We iterate the forward process and the error back-propagation by a few million times, and record the final error $E$ after equilibrium is reached. Various size of hidden layers have been examined, and we have found} \vp{the} minimum $H$ for a successful storage of patterns with an accuracy \pa{of} $E<2^{-N}$. We have considered \jj{more than $5000$} ensembles starting from different initial parameter values, because the nonlinear feedforward equations have many local minima in the error landscape. \paNew{Although BP algorithm is an old technique, its use in this study merits a direct relation between the network size and the complexity of the input signal patterns. Additionally, the extensive use of BP~\citep{Watkin1993, Rumelhart1986, LeCun1998} and its various hybrid forms (modified BP or combination with other learning algorithms), imply that the minimal perceptrons to be found may give insights on the first principles of network size optimization for memory storage based on the complexity of input signals.}

\begin{figure}
\begin{center}
\includegraphics[width=100mm]{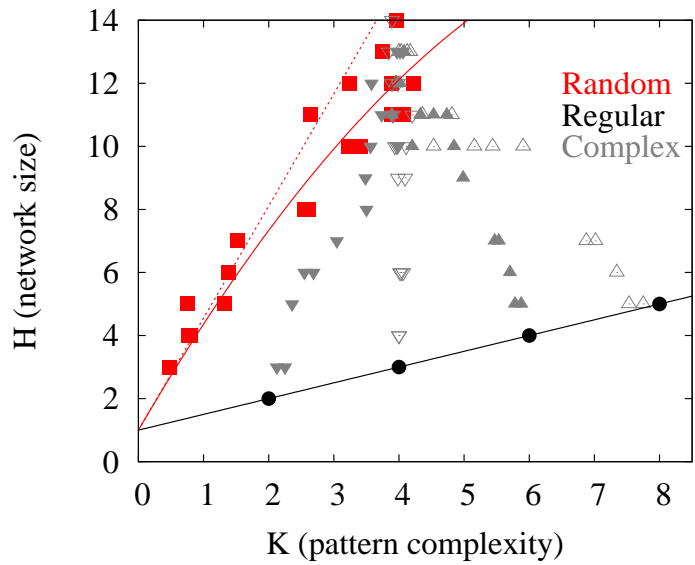}
\caption{(Color online) Pattern complexity and network size.
Random (red square), regular (black circle), and complex binary patterns (gray triangles)
with an input size $N=8$ are used.
The complex patterns are generated by shuffling elements in
the regular patterns of
(8, 6), (8, 4), (8, 2), and (8, 0) pseudo-parity patterns 
(lower filled, lower empty, upper filled, and upper empty triangles, respectively).
The lines are theoretical estimations for regular patterns [black, Eq.~(\ref{eq:H1})]
and for random patterns [dotted red, Eq~\ref{eq:H2a}, and solid red, Eq.~(\ref{eq:H2b})].
\label{fig4}
}
\end{center}
\end{figure}

As shown in Fig.~\ref{fig4}, the minimum $H$ obtained in the simulation is consistent with the estimations \pa{of} Eq.~(\ref{eq:H1}) for regular patterns and Eq.~(\ref{eq:H2b}) for random patterns.



\section{Minimal network for complex patterns} \label{sec:complex}
The strategies for memorizing regular and random patterns can be
applied to memorize complex patterns, which are neither completely ordered nor disordered.
The minimal perceptrons for describing complex patterns \vp{should} require
$H$ between $H_1$ in Eq.~(\ref{eq:H1}) for regular patterns 
and $H_2$ in Eq.~(\ref{eq:H2b}) for random patterns (Fig.~\ref{fig4}):
\begin{equation} \label{eq:H3a}
H = (1-\lambda) H_1 + \lambda H_2
\end{equation}
with $0 < \lambda < 1$.
A complex pattern $\bm \xi$ can be decomposed into ordered elements, projected on a regular template pattern $\bm \xi_1$,
and disordered elements, impurities deviated from the template.
The Hamming distance between $\bm \xi$ and $\bm \xi_1$ gives the number of impurities:
$D=\sum_{\mu=1}^M (1- \delta_{z^\mu, z_1^\mu})$.
Here the impurity fraction is defined as $d\equiv D/M$.
The dichotomy for the complex pattern uses
$H_1$ hidden nodes for the regular elements and additional $\Delta H$ hidden nodes
for the impurity elements.
Then, the activation of \pa{the} output \vp{node} becomes $\sum_{j=1}^{H_1+\Delta H} v_j y_j -v_0$.
Here the additional hidden nodes reverse the reference activation $\sum_{j=1}^{H_1} v_j y_j -v_0$ for the regular elements
by adding a larger positive or negative value only for the impurity elements.
The process \pa{of} distinguishing $dM$ impurity elements from $(1-d)M$ regular ones is
the same \vp{as} the process \pa{of} distinguishing $pM$ black elements  from $(1-p)M$ white ones.
Thus the formula, $H_2[K(p)]$ in Eq.~(\ref{eq:H2b}), can be applied to estimate $\Delta H = H_2(d)$.
\begin{figure}
\begin{center}
\includegraphics[width=100mm]{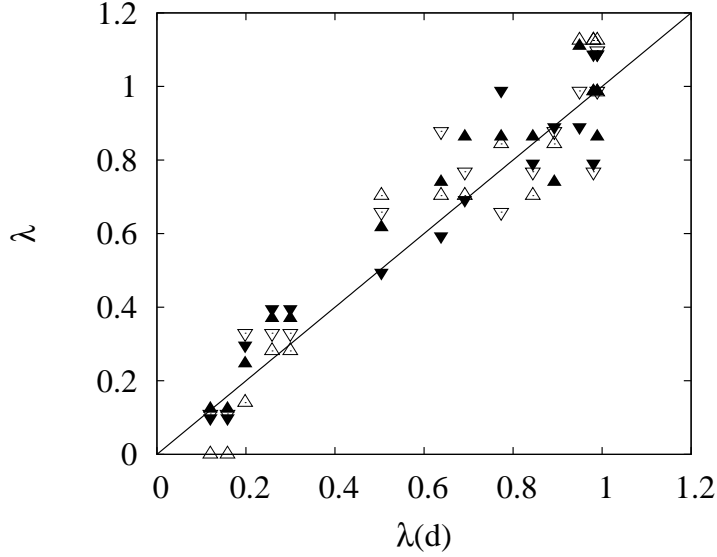}
\caption{Interpolation of complex patterns.
The interpolation parameter $\lambda$
is compared with the predicted $\lambda(d)$ based on the
Hamming distance $d$ between complex patterns and
regular patterns. 
The complex patterns are generated by shuffling elements in
the regular patterns of
(8, 6), (8, 4), (8, 2), and (8, 0) pseudo-parity patterns 
(lower filled, lower empty, upper filled, and upper empty triangles, respectively).
\label{fig5}
}
\end{center}
\end{figure}

\pa{In the absence of impurity} ($d=0$), the pattern corresponds to a regular pattern ($H=H_1$).
On the other hand, when the impurity is \vp{maximal} ($d=p$),
the required number of hidden nodes should become $H=H_2(p)$.
This \pa{constraint} introduces a scale factor $[{H_2(p)-H_1}]/{H_2(p)}$ for $\Delta H$.
Therefore, the minimum $H$ for complex patterns is
\begin{equation} \label{eq:H3b}
H = H_1 + H_2(d) \cdot \frac{H_2(p)-H_1}{H_2(p)}.
\end{equation}
The comparison of this equation with Eq.~(\ref{eq:H3a}) gives the linear factor,
\begin{equation} \label{eq:lambda}
\lambda(d) = \frac{H_2(d)}{H_2(p)}.
\end{equation}
Indeed, the Hamming distance $d$ \vp{can} be used to predict the linear factor $\lambda$ (Fig.~\ref{fig5})
\pa{and allows} to estimate $H$ in Eq.~(\ref{eq:H3a}) for complex patterns.

\section{Conclusion}
\label{sec:conclusion}
\pa{Perceptrons} encode a pattern into their weights and biases,
\vp{which amounts to} data compression in information theory~\citep{Cover2006}.
In machine learning, determining the
appropriate \vp{size} of perceptrons to encode patterns is an old unsolved problem.
Here we introduced a simple complexity index \vp{for} patterns,
and derived \vp{a} minimum number of hidden nodes, dependent on \vp{this} complexity index,
for memorizing patterns.
However, further study is needed to examine 
how \vp{additional nodes}  and complexities of network structure 
affect the robustness and adaptability of pattern storage.

\section{Acknowledgements}
We thank V. Periwal for discussions and hospitality during our stay at National Institutes of Health in the US. \paNew{M. Pastor acknowledges the Office of Research, University of San Carlos, Cebu City, Philippines.}
This research was supported by Basic Science Research Program
through the National Foundation of Korea funded by the Ministry
of Science, ICT \& Future Planning (No. 2013R1A1A1006655) 
and by the Max Planck Society, the Korea Ministry of Education, Science and Technology,
Gyeongsangbuk-Do and Pohang City.


\begin{thebibliography}{23}
\providecommand{\natexlab}[1]{#1}
\providecommand{\url}[1]{\texttt{#1}}
\providecommand{\href}[2]{#2}
\providecommand{\path}[1]{#1}
\providecommand{\eprint}[1]{\href{http://arxiv.org/abs/#1}{\path{#1}}}
\providecommand{\DOIprefix}{doi:}
\providecommand{\ArXivprefix}{arXiv:}
\providecommand{\URLprefix}{URL: }
\providecommand{\Pubmedprefix}{pmid:}
\providecommand{\doi}[1]{\href{http://dx.doi.org/#1}{\path{#1}}}
\providecommand{\Pubmed}[1]{\href{pmid:#1}{\path{#1}}}
\providecommand{\BIBand}{and}
\providecommand{\bibinfo}[2]{#2}
\ifx\xfnm\undefined \def\xfnm[#1]{\unskip,\space#1}\fi
\bibitem[{Ou and Murphey(2007)}]{GOu2007}
\bibinfo{author}{Ou\xfnm[ G.]}, \bibinfo{author}{Murphey\xfnm[ Y.L.]}.
\newblock \bibinfo{title}{Multi-class pattern classification using neural
  networks}.
\newblock \bibinfo{journal}{Pattern Recognition}
  \bibinfo{year}{2007};\bibinfo{volume}{40}:\bibinfo{pages}{4--18}.
\bibitem[{Monterola and Zapotocky(2005)}]{Monterola2005}
\bibinfo{author}{Monterola\xfnm[ C.]}, \bibinfo{author}{Zapotocky\xfnm[ M.]}.
\newblock \bibinfo{title}{Noise-enhanced categorization in recurrently
  reconnected neural network}.
\newblock \bibinfo{journal}{Physical Review E}
  \bibinfo{year}{2005};\bibinfo{volume}{71}:\bibinfo{pages}{036164}.
\newblock \DOIprefix\doi{10.1103/PhysRevE.71.036134}.
\bibitem[{Hopfield(1982)}]{Hopfield1982}
\bibinfo{author}{Hopfield\xfnm[ J.]}.
\newblock \bibinfo{title}{Neural networks and physical systems with emergent
  collective computational abilities}.
\newblock \bibinfo{journal}{Proceedings of National Academy of Sciences}
  \bibinfo{year}{1982};\bibinfo{volume}{79}:\bibinfo{pages}{2554--2558}.
\bibitem[{Abbott and Arian(1987)}]{Abbott1987}
\bibinfo{author}{Abbott\xfnm[ L.]}, \bibinfo{author}{Arian\xfnm[ Y.]}.
\newblock \bibinfo{title}{Storage capacity of generalized networks}.
\newblock \bibinfo{journal}{Physical Review A}
  \bibinfo{year}{1987};\bibinfo{volume}{36}:\bibinfo{pages}{5091}.
\newblock \DOIprefix\doi{10.1103/PhysRevA.36.5091}.
\bibitem[{Shiino and Fukai(1993)}]{Shiino1993}
\bibinfo{author}{Shiino\xfnm[ M.]}, \bibinfo{author}{Fukai\xfnm[ T.]}.
\newblock \bibinfo{title}{Self-consistent signal-to-noise analysis of the
  statistical behavior of analog neural networks and enhancement of the storage
  capacity}.
\newblock \bibinfo{journal}{Physical Review E}
  \bibinfo{year}{1993};\bibinfo{volume}{48}:\bibinfo{pages}{867}.
\newblock \DOIprefix\doi{10.1103/PhysRevE.48.867}.
\bibitem[{Watkin and Rau(1993)}]{Watkin1993}
\bibinfo{author}{Watkin\xfnm[ T.]}, \bibinfo{author}{Rau\xfnm[ A.]}.
\newblock \bibinfo{title}{The statistical mechanics of learning a rule}.
\newblock \bibinfo{journal}{Reviews of Modern Physics}
  \bibinfo{year}{1993};\bibinfo{volume}{65}:\bibinfo{pages}{499--556}.
\newblock \DOIprefix\doi{10.1103/RevModPhys.65.499}.
\bibitem[{Blum and Rivest(1992)}]{Blum1992}
\bibinfo{author}{Blum\xfnm[ A.]}, \bibinfo{author}{Rivest\xfnm[ R.]}.
\newblock \bibinfo{title}{Training a 3-node neural network is np-complete}.
\newblock \bibinfo{journal}{Neural Networks}
  \bibinfo{year}{1992};\bibinfo{volume}{5}:\bibinfo{pages}{117--127}.
\newblock \DOIprefix\doi{10.1016/S0893-6080(05)80010-3}.
\bibitem[{Murata~N. and S.I.(1994)}]{Murata1994}
\bibinfo{author}{Murata~N.\xfnm[ Y.S.]}, \bibinfo{author}{S.I.\xfnm[ A.]}.
\newblock \bibinfo{title}{Network information criterion-determining the number
  of hidden units for an artificial neural network model}.
\newblock \bibinfo{journal}{IEEE Transactions on Neural Networks}
  \bibinfo{year}{1994};\bibinfo{volume}{5}:\bibinfo{pages}{865--872}.
\newblock \DOIprefix\doi{10.1109/72.329683}.
\bibitem[{Hirose~Y. and S.(1991)}]{Hirose1991}
\bibinfo{author}{Hirose~Y.\xfnm[ Y.K.]}, \bibinfo{author}{S.\xfnm[ H.]}.
\newblock \bibinfo{title}{Back-propagation algorithm which varies the number of
  hidden units}.
\newblock \bibinfo{journal}{Neural Networks}
  \bibinfo{year}{1991};\bibinfo{volume}{4}:\bibinfo{pages}{61--66}.
\newblock \DOIprefix\doi{10.1016/0893-6080(91)90032-Z}.
\bibitem[{Hunter~D. and B.(2012)}]{Hunter2012}
\bibinfo{author}{Hunter~D. Yu~H.\xfnm[ P.M.K.J.]}, \bibinfo{author}{B.\xfnm[
  W.]}.
\newblock \bibinfo{title}{Selection of proper neural network sizes and
  architectures - a comparative study}.
\newblock \bibinfo{journal}{IEEE Transactions on Industrial Information}
  \bibinfo{year}{2012};\bibinfo{volume}{8}:\bibinfo{pages}{228--240}.
\newblock \DOIprefix\doi{10.1109/TII.2012.2187914}.
\bibitem[{Kwok and Yeung(1997)}]{Kwok1997}
\bibinfo{author}{Kwok\xfnm[ T.]}, \bibinfo{author}{Yeung\xfnm[ D.]}.
\newblock \bibinfo{title}{Constructive algorithms for structure learning in
  feedforward neural networks for regression problems}.
\newblock \bibinfo{journal}{IEEE Transactions on Neural Networks}
  \bibinfo{year}{1997};\bibinfo{volume}{8}:\bibinfo{pages}{630--645}.
\newblock \DOIprefix\doi{10.1109/72.572102}.
\bibitem[{Reed(1993)}]{Reed1993}
\bibinfo{author}{Reed\xfnm[ R.]}.
\newblock \bibinfo{title}{Pruning algorithms-a survey}.
\newblock \bibinfo{journal}{IEEE Transactions on Neural Networks}
  \bibinfo{year}{1993};\bibinfo{volume}{4}:\bibinfo{pages}{740--747}.
\newblock \DOIprefix\doi{10.1109/72.248452}.
\bibitem[{Castellano~G. and M.(1997)}]{Castellano1997}
\bibinfo{author}{Castellano~G.\xfnm[ F.A.]}, \bibinfo{author}{M.\xfnm[ P.]}.
\newblock \bibinfo{title}{An iterative pruning algorithm for feedforward neural
  networks}.
\newblock \bibinfo{journal}{IEEE Transactions on Neural Networks}
  \bibinfo{year}{1997};\bibinfo{volume}{8}:\bibinfo{pages}{519--531}.
\newblock \DOIprefix\doi{10.1109/72.572092}.
\bibitem[{Shannon(1948)}]{Shannon1948}
\bibinfo{author}{Shannon\xfnm[ C.]}.
\newblock \bibinfo{title}{A mathematical theory of communication}.
\newblock \bibinfo{journal}{Bell System Technical Journal}
  \bibinfo{year}{1948};\bibinfo{volume}{27}:\bibinfo{pages}{379--423}.
\newblock \DOIprefix\doi{10.1002/j.1538-7305.1948.tb01338.x}.
\bibitem[{Silva(2015)}]{Silva2015}
\bibinfo{author}{Silva Luiz Eduardo~Virgilio\xfnm[ e.a.]}.
\newblock \bibinfo{title}{Multiscale entropy-based methods for heart rate
  variability complexity analysis}.
\newblock \bibinfo{journal}{Physica A: Statistical Mechanics and its
  Applications}
  \bibinfo{year}{2015};\bibinfo{volume}{422}:\bibinfo{pages}{143--152}.
\newblock \DOIprefix\doi{10.1016/j.physa.2014.12.011}.
\bibitem[{Larrondo(2005)}]{Larrondo2005}
\bibinfo{author}{Larrondo H.A\xfnm[ e.a.]}.
\newblock \bibinfo{title}{Intensive statistical complexity measure of
  pseudorandom number generators}.
\newblock \bibinfo{journal}{Physica A: Statistical Mechanics and its
  Applications}
  \bibinfo{year}{2005};\bibinfo{volume}{356.1}:\bibinfo{pages}{133--138}.
\newblock \DOIprefix\doi{10.1016/j.physa.2005.05.025}.
\bibitem[{Cover and Thomas(2006)}]{Cover2006}
\bibinfo{author}{Cover\xfnm[ T.]}, \bibinfo{author}{Thomas\xfnm[ J.]}.
\newblock \bibinfo{title}{Elements of information theory}.
\newblock \bibinfo{publisher}{Wiley-Interscience}; \bibinfo{year}{2006},.
\bibitem[{LeCun~Y. and K.R.(1998)}]{LeCun1998}
\bibinfo{author}{LeCun~Y. Bottou~L.\xfnm[ O.G.]}, \bibinfo{author}{K.R.\xfnm[
  M.]}.
\newblock \bibinfo{title}{Efficient backdrop}.
\newblock \bibinfo{publisher}{Springer Berlin Heidelberg};
  \bibinfo{year}{1998},.
\bibitem[{Haykin(1999)}]{Haykin1999}
\bibinfo{author}{Haykin\xfnm[ S.]}.
\newblock \bibinfo{title}{Neural networks: a comprehensive foundation}.
\newblock \bibinfo{publisher}{New York Prentice-Hall, Inc.};
  \bibinfo{year}{1999},.
\bibitem[{Cybenko(1989)}]{Cybenko1989}
\bibinfo{author}{Cybenko\xfnm[ G.]}.
\newblock \bibinfo{title}{Approximation by superpositions of a sigmoidal
  function}.
\newblock \bibinfo{journal}{Mathematics of Control Signals, and Systems}
  \bibinfo{year}{1989};\bibinfo{volume}{2}:\bibinfo{pages}{303--314}.
\newblock \DOIprefix\doi{10.1007/BF02551274}.
\bibitem[{Setiono(1997)}]{Setiono1997}
\bibinfo{author}{Setiono\xfnm[ R.]}.
\newblock \bibinfo{title}{On the solution of the parity problem by a single
  hidden layer feedforward neural network}.
\newblock \bibinfo{journal}{Neurocomputing}
  \bibinfo{year}{1997};\bibinfo{volume}{16}:\bibinfo{pages}{225--235}.
\newblock \DOIprefix\doi{10.1016/S0925-2312(97)00030-1}.
\bibitem[{Sontag(1992)}]{Sontag1992}
\bibinfo{author}{Sontag\xfnm[ E.]}.
\newblock \bibinfo{title}{Feedforward nets for interpolation and
  classification}.
\newblock \bibinfo{journal}{Journal of Computer and System Sciences}
  \bibinfo{year}{1992};\bibinfo{volume}{45}:\bibinfo{pages}{20--48}.
\newblock \DOIprefix\doi{10.1016/0022-0000(92)90039-L}.
\bibitem[{Rumelhart~D. and R.(1986)}]{Rumelhart1986}
\bibinfo{author}{Rumelhart~D.\xfnm[ H.G.]}, \bibinfo{author}{R.\xfnm[ W.]}.
\newblock \bibinfo{title}{Learning representations by back-propagating errors}.
\newblock \bibinfo{journal}{Nature}
  \bibinfo{year}{1986};\bibinfo{volume}{323}:\bibinfo{pages}{533--536}.

\end{thebibliography}
\end{document}